\input harvmac.tex
\def\ap{\alpha^\prime}
\let\includefigures=\iftrue
\newfam\black
\includefigures
\input epsf
\def\figin{\epsfcheck\figin}\def\figins{\epsfcheck\figins}
\def\epsfcheck{\ifx\epsfbox\UnDeFiNeD
\message{(NO epsf.tex, FIGURES WILL BE IGNORED)}
\gdef\figin##1{\vskip2in}\gdef\figins##1{\hskip.5in}
\else\message{(FIGURES WILL BE INCLUDED)}%
\gdef\figin##1{##1}\gdef\figins##1{##1}\fi}
\def\P{\BP}
\def\BP{\IP}
\def\IP{\relax{\rm I\kern-.18em P}}

\def\IR{\relax{\rm I\kern-.18em R}}
\def\DefWarn#1{}
\def\figinsert{\goodbreak\midinsert}
\def\ifig#1#2#3{\DefWarn#1\xdef#1{fig.~\the\figno}
\writedef{#1\leftbracket fig.\noexpand~\the\figno}%
\figinsert\figin{\centerline{#3}}\medskip\centerline{\vbox{\baselineskip12pt
\advance\hsize by -1truein\noindent\footnotefont{\bf Fig.~\the\figno:} #2}}
\bigskip\endinsert\global\advance\figno by1}
\else
\def\ifig#1#2#3{\xdef#1{fig.~\the\figno}
\writedef{#1\leftbracket fig.\noexpand~\the\figno}%
\global\advance\figno by1} 
\fi

\Title{\vbox{\baselineskip12pt\hbox{hep-th/0103115}
\hbox{CALT-68-2320}\hbox{CITUSC/01-004}}}
{\vbox{
\centerline{D-Branes, Holonomy and M-Theory}
\vskip 8pt
}}
\centerline{Jaume Gomis}
\medskip
\medskip
\medskip
\vskip 8pt 
\centerline{\it Department of Physics}
\centerline{\it California Institute of Technology}
\centerline{\it Pasadena, CA 91125}
\centerline{\it and}
\centerline{\it Caltech-USC Center for Theoretical Physics} 
\centerline{\it University of Southern California}
\centerline{\it Los Angeles, CA 90089}

\medskip
\centerline{}
\medskip
\medskip
\medskip
\noindent

We show that M-theory on spaces with irreducible holonomy 
represent Type IIA backgrounds in which a collection of D6-branes wrap
a supersymmetric cycle in a manifold with a  holonomy group different
from the one  appearing in the M-theory description. For example, we
show that D6-branes wrapping a supersymmetric four-cycle 
 on a manifold with $G_2$ holonomy is described in eleven dimensions
by M-theory  on a 
space with $\hbox{Spin}(7)$ holonomy. Examples of such Type IIA
backgrounds which lift to M-theory on spaces with $SU(3), G_2, SU(4)$
and $\hbox{Spin}(7)$ holonomy are considered. The M-theory geometry
can then be used to compute exact quantities of the gauge theory
on the corresponding D-brane configuration.

\smallskip

\Date{March 2001}

\newsec{Introduction}

One of the important lessons of the duality era has been that string
theory backgrounds can be naturally embedded in an eleven dimensional
theory. The simplest example of this embedding is the uplift of
any background of Type IIA string theory to M-theory
\nref\townsend{P.K. Townsend, ``The eleven-dimensional supermembrane
revisited'', Phys.Lett. {\bf B350} (1995) 184.}%
\nref\witten{E. Witten, ``String Theory Dynamics In Various Dimensions'',
Nucl.Phys. {\bf B443} (1995) 85.}%
\refs{\townsend,\witten}. In particular,
any solution of the equations of motion described by the
ten dimensional metric $g_{\mu\nu}$, the Ramond-Ramond one form $C_\mu$
and the dilaton $\phi$ uplifts in eleven dimensions to a solution of eleven
dimensional supergravity with the following eleven dimensional metric
\eqn\metri{
ds_{11}^2=e^{-{2\phi\over 3}}g_{\mu\nu}dx^\mu dx^\nu +e^{{4\phi\over
3}}(dx_{11}+C_{\mu} dx^{\mu})^2.}
The uplift of the rest of the massless bosonic fields of Type IIA string
theory corresponds to turning on the three-form gauge field of eleven
dimensional supergravity. Therefore, backgrounds of Type IIA string
theory  which only  
source the fields $g_{\mu\nu}$, $C_\mu$ and $\phi$ must be described in
eleven dimensions by a purely gravitational background without any
flux.

On the other hand, supersymmetric M-theory vacua which are purely
gravitational are completely 
classified. Supersymmetric compactifications of
eleven dimensional 
supergravity to $d+1$-dimensional Minkowski space ${\bf R}^{1,d}$
appear whenever   the
$D\equiv 10-d$  
dimensional compactification manifold ${\bf X}$ admits covariantly 
constant spinors\foot{There are many
other interesting supersymmetric solutions if one allows for
non-trivial background 
fluxes. For example, one can have AdS vacua (see 
\nref\ads{O.Aharony, S.S. Gubser, J. Maldacena, H. Ooguri and Y.Oz, ``Large N
Field Theories, String Theory and Gravity'',  
Phys.Rept. {\bf 323} (2000) 183; M.J. Duff, ``TASI Lectures on Branes,
Black Holes and Anti-de Sitter  
Space'', hep-th/9912164.}%
\ads\ for a nice review of some such solutions) and warped
compactifications (see e.g. 
\nref\strominger{A. Strominger, ``Superstrings with Torsion'',
Nucl.Phys. {\bf B274} (1986) 253.}%
\nref\beckers{K. Becker and M. Becker, ``M-Theory on
Eight-Manifolds'', Nucl.Phys. {\bf B477} (1996) 155.}%
\refs{\strominger,\beckers}).}. The only solution in
this class with maximal supersymmetry 
is when ${\bf X}$ is a torus ${\bf T}^{D}$. Solutions with reduced
supersymmetry can be obtained when the compactification space ${\bf X}$ has
 covariantly constant spinors which do not span the full $SO(D)$
spinor. Such $D$-dimensional spaces have a holonomy group $G$ such
that at least one component of the $SO(D)$ spinor is left invariant by
the action of $G$.
The list 
\nref\berger{M. Berger, ``Sur les groupes d'holonomie homog\`ene de
vari\'et\'es \`a connexion affines et des vari\'et\'es
riemanniennes'', Bull. Soc. Math. France {\bf 83} (1955) 279.}%
\berger\ of irreducible holonomy groups preserving some supersymmetry 
and the corresponding number of unbroken
supercharges in ${\bf R}^{1,d}$ is : 
\medskip
\noindent
$\bullet\ D=4, \quad G=SU(2),\quad 7\ \hbox{dim.}\ {\cal N}=1\
\hbox{SUSY}, \qquad 16\ 
\hbox{real supercharges}$ 

\noindent
$\bullet\ D=6, \quad G=SU(3), \quad 5\ \hbox{dim.}\ {\cal N}=1\
\hbox{SUSY},\qquad 8\ \hbox{real supercharges}$

\noindent
$\bullet\ D=7, \quad G=G_2, \quad \hskip+15pt 4\ \hbox{dim.}\ {\cal N}=1\
\hbox{SUSY}, \qquad 4\ \hbox{real supercharges}$

\noindent
$\bullet\ D=8, \quad G=SU(4), \quad  3\ \hbox{dim.}\ {\cal N}=2\
\hbox{SUSY}, \qquad 4\ \hbox{real supercharges}$

\noindent
$\bullet\ D=8, \quad G=\hbox{Spin}(7), \quad 3\ \hbox{dim.}\ {\cal N}=1\
\hbox{SUSY}, \qquad 2\ \hbox{real supercharges}$

\medskip 

The number of unbroken supercharges is given by the number of singlets
in the decomposition of the spinor of $SO(D)$ under the
reduced holonomy group $G \subset \hbox{Spin}(D)$.
Manifolds with $SU(2), SU(3)$ and
$SU(4)$ holonomy are Calabi-Yau spaces while manifolds with $G_2$ and
$\hbox{Spin}(7)$ holonomy are real Ricci flat
manifolds\foot{Eight dimensional hyper-Kahler manifolds with 
 $Sp(2)$ holonomy will not be considered in this paper. They give rise
to three dimensional theories with ${\cal N}=3$ supersymmetry.}.

In this paper we show that  M-theory
compactified\foot{We will consider non-compact spaces
so strictly speaking we are not compactifying. Therefore, the constraint
\nref\sevawi{S. Sethi, C. Vafa and E. Witten, ``Constraints on
Low-Dimensional String Compactifications'', Nucl.Phys. {\bf B480}
(1996) 213, hep-th/9606122.}%
\sevawi\ imposed by tadpole cancellation is lifted since the flux lines
can escape to infinity.}
on certain spaces with reduced holonomy appear as  the local eleven
dimensional 
description of Type IIA D6-branes wrapping
supersymmetric cycles on certain lower dimensional spaces with
a different holonomy group than the one in the M-theory
description. It is crucial that we use D6-branes in the Type IIA
construction since this is the only\foot{The magnetic dual object, the
D0-brane, also lifts to pure geometry.} brane which lifts in eleven
dimensions to pure geometry.  Supersymmetry severely constraints the
types of lifts that are possible.
The basic geometry that appears in eleven
dimensions near the location of the D6-branes is that of an ALE
fibration over the supersymmetric cycle 
in which we wrapped the D6-branes. This fibration has a different
holonomy group than that of the manifold we started with in Type IIA.
We consider the possibility of
adding curved orientifold  six planes (O6-planes) by modding out by
worldsheet parity 
combined with an appropriate involution on the geometry. Then, the type of
ALE fibration appearing in eleven dimensions depends on whether there
is an orientifold plane in the Type IIA description. The gauge theory
on the D6-branes appears in the M-theory description from
singularities in the geometry. In the various examples we consider, we
notice that the M-theory geometrical description can be used to
compute quantities -- such as prepotentials or  superpotentials -- of the
gauge theory living on the D6-branes.

Our results extend the work of
\nref\acha{B.S. Acharya, ``On Realising N=1 Super Yang-Mills in M
theory'', hep-th/0011089.}%
\nref\atmalva{M. Atiyah, J. Maldacena and C. Vafa, ``An M-theory Flop
as a Large N Duality'', hep-th/0011256.}%
\acha\atmalva\ where a configuration of D6 branes wrapping the ${\bf S}^3$
which appears in the deformation of the conifold singularity of a
Calabi-Yau three-fold was represented in M-theory as 
compactification on a certain space with $G_2$ holonomy.
For example, we show that M-theory on an orbifold of the spin
bundle over ${\bf S}^4$  $S({\bf S}^4)$ -- which admits a metric with
$\hbox{Spin}(7)$ holonomy -- appears 
as the strong coupling description of D6-branes wrapping the
supersymmetric four-cycle of the bundle of anti-self-dual two-forms
over ${\bf S}^4$ $\Sigma({\bf S}^4)$ which admits a $G_2$ holonomy metric. 
 This type of correspondence is extended to all
other cases. Namely, we show that M-theory on  certain singular
spaces with $SU(3), G_2, SU(4)$ and   $\hbox{Spin}(7)$ holonomy
describe in eleven dimensions a configuration of D6-branes wrapped on
a supersymmetric cycle on a manifold with different holonomy
group. For example, we show that  M-theory on the
Calabi-Yau three-folds used in 
\nref\geom{S. Katz, P. Mayr and C. Vafa, ``Mirror symmetry and Exact
Solution of 4D N=2 Gauge Theories I'', Adv.Theor.Math.Phys. {\bf 1}
(1998) 53.}%
\geom\
 to  geometrically  engineer
supersymmetric  gauge theories 
theories appear as the local description of D6-branes wrapping two-cycles
in manifolds with $SU(2)$ holonomy.
We would
like to emphasize that the M-theory descriptions that we propose in
this paper are appropriate descriptions of the Type IIA brane
configurations near the location of the D6-branes. Globally, the
eleven dimensional geometry is more complicated. 
In this paper we will provide  simple examples of the local
description of the wrapped supersymmetric D6-branes configurations in
terms of eleven dimensional geometry.

This phenomena, apart from giving a physical rational for the
existence of manifolds with reduced holonomy, can be useful in
deriving  string theory dualities using geometrical transitions. Recently,
Atiyah, Maldacena and Vafa \atmalva\ have lifted the description of Type IIA
backgrounds on two different  Calabi-Yau three-folds to eleven
dimensions and have found that the two different Type IIA backgrounds
are described in M-theory by two different $G_2$ holonomy spaces which 
are related to each other by a flop transition \atmalva . Therefore, by going
through a smooth physical transition in M-theory, like a  flop
in a $G_2$ holonomy manifold,
they derive 
the physical equivalence of string theory in the two different
Calabi-Yau manifolds, which was previously conjectured by Vafa
\nref\vafa{C. Vafa, ``Superstrings and Topological Strings at Large
N'', hep-th/0008142.}%
\vafa . Various aspects of this duality and generalizations to other
three-folds have recently appeared in 
\nref\sinvafa{S. Sinha and C.Vafa, ``SO and Sp Chern-Simons at Large
N'', hep-th/0012136.}%
\nref\achar{B.S. Acharya, ``Confining Strings from $G_2$-holonomy
spacetimes'', hep-th/0101206.}%
\nref\achavafa{B.S. Acharya and C. Vafa, ``On Domain Walls of N=1
Supersymmetric Yang-Mills in Four Dimensions'', hep-th/0103011.}%
\nref\cacintrivafa{F. Cachazo, K. Intriligator and C. Vafa, ``A Large
N Duality via a Geometric Transition'', hep-th/0103067.}%
\refs{\sinvafa,\achar,\achavafa,\cacintrivafa}. The lifts proposed in
this paper  provide quantitative information about the gauge
theory on the branes and can be useful in deriving possible new dualities among
Type IIA vacua by going through a geometrical transition in the M-theory
geometries we discuss.

The rest of the paper is organized as follows. Section $2$ briefly
reviews the eleven dimensional description of Type IIA D6-branes in
flat space with and
without an orientifold six-plane.
In section $3$ we discuss the gauge theories one gets by wrapping
D6-branes on supersymmetric cycles. We also exhibit the general features of
the eleven dimensional geometry that represents the various wrapped
D6-brane configurations. Sections
$4$-$7$ give specific examples  of lifts of Type IIA backgrounds
with branes on 
certain manifolds with one holonomy group in terms of M-theory
on spaces with a different holonomy group.

\newsec{The Geometry of D6-branes and the O6-Plane}

Any solution of the Type IIA string equations of motion only sourcing  
the Type IIA fields $g_{\mu\nu}$, $C_\mu$ and $\phi$ is described in
eleven dimensions by a purely gravitational background. A simple
example of this  phenomena is provided by a collection of
D6-branes. The eleven
dimensional description of N separated D6-branes can be obtained by using
\metri\ on the Type IIA supergravity solution of 
\nref\horostro{G.T. Horowitz and A. Strominger, ``Black Strings and
P-Branes'', Nucl.Phys. {\bf B360} (1991) 197.}%
\horostro . It is given by  
\eqn\lift{
ds_{11}^2=-dx_0^2+dx_1^2+\ldots++dx_6^2+ds_{TN}^2,}
where $ds_{TN}^2$ is the metric of the Euclidean multi-centered
Taub-NUT space
\nref\gibbhaw{S. Hawking, ``Gravitational Instantons'',
Phys.Lett. {\bf A60} (1977) 81; G.W. Gibbons and S.W. Hawking,
``Gravitational Multi-Instantons'',  
Phys.Lett. {\bf B78} (1978) 430.}
\gibbhaw
\eqn\taub{\eqalign{
ds_{TN}^2&=H d\vec{r}^{\;2}+H^{-1}(dx_{11}+C_{\mu} dx^{\mu})^2\cr
\vec{\nabla}\times \vec{C} &=-\vec{\nabla}H\cr
H&=1+{1\over 2}\sum_{i=1}^N{g_s \sqrt{{\ap}}\over |\vec{r}-\vec{r_i}|},}}
which is a hyper-Kahler metric on a $U(1)$  bundle over ${\bf R}^3$. 
 The $x^{11}$
coordinate is periodic and absence of conical
singularities  at ${\vec r}={\vec r}_i$ requires
its periodicity to be $2\pi g_s\sqrt{\ap}$.
The circle fiber
is identified with the M-theory circle which has the expected radius
$R=g_s\sqrt{\ap}$ at infinity.   
Therefore,  a collection of D6-branes in flat space can be represented in
M-theory by a four dimensional  manifold
with $SU(2)$ holonomy.

In this paper we take the D6-branes to be coincident --  all
$\vec{r_i}=0$ --  which gives rise to the familiar enhanced $SU(N)$ gauge
symmetry. The M-theory description of this configuration near
the location of the D6-branes at ${\vec r}\simeq 0$ is given by
\eqn\ale{\eqalign{
ds^2_{ALE}&\simeq H d\vec{r}^{\;2}+H^{-1}(dx_{11}+C_{\mu} dx^{\mu})^2\cr
H&\simeq{Ng_s \sqrt{{\ap}}\over 2r},}}
which is the metric on the $A_{N-1}$ ALE space 
\nref\eguchihan{T. Eguchi and A.J. Hanson, ``Asymptotically Flat
Self-Dual Solutions to Euclidean Gravity'', Phys.Lett. {bf 74B} (1978)
249; T.Eguchi, P.B. Gilkey and A.J. Hanson, ``Gravitation, Gauge
Theories and Differential Geometry'', Phys.Rep. {\bf 66} (1980) 214.}%
\eguchihan\ near an  orbifold singularity. Here the $SU(N)$ gauge
symmetry arises from membranes wrapping the collapsed two-cycles
\witten . In the limit when all $\vec{r_i}=0$
the metric on the $A_{N-1}$ ALE space \eguchihan\  degenerates to the
metric on the  
${\bf C}^2/{\bf Z}_N$ orbifold, where ${\bf Z}_N$ acts by
\eqn\znact{\eqalign{
z_1&\rightarrow e^{2\pi i\over N}z_1\cr
z_2&\rightarrow e^{-{2\pi i\over N}}z_2.}}
Three of the coordinates which are acted on are  the original
transverse directions of the D6 branes. The fourth orbifolded
direction corresponds to the M-theory circle whose asymptotic radius
is now infinity. Therefore, Type IIA string theory with $N$ coincident
D6-branes  
is locally described in
eleven dimensions by the $A_{N-1}$ ALE singularity.

One can also find the  M-theory realization of a
collection of D6-branes sitting on top of an orientifold six-plane
(O6-plane). The O6-plane appears as the fixed locus obtained by
modding out string theory by the orientifold group $G=\{1,\Omega
(-1)^{F_L}R_{7}R_{8}R_{9}\}$, where
 $\Omega$ is 
worldsheet parity,
${F_L}$ is the left-moving space-time fermion number and $R_i$ is a
reflection along the $x^i$-th coordinate. In this paper we 
consider M-theory lifts involving the O6-plane which carries -2 units of
D6-brane charge\foot{In this paper all charges are
 measured in the quotient space.} (the O6$^-$ plane). $N$
 coincident D6-branes on top of this
orientifold plane gives rise to enhanced $SO(2N)$ gauge symmetry.
It was shown in 
\nref\seiberg{N. Seiberg, ``IR Dynamics on Branes and Space-Time
Geometry'', Phys.Lett. {\bf B384} (1996) 81.}%
\nref\seibwit{N. Seiberg and E. Witten, ``Gauge Dynamics And
Compactification To 
Three Dimensions'', hep-th/9607163.}%
\refs{\seiberg,\seibwit} that this O6-plane is represented in M-theory
by the Atiyah-Hitchin space 
\nref\atiyahhitch{M.F. Atiyah and N.J. Hitchin, ``Low-Energy
Scattering of Nonabelian Monopoles'', Phys. Lett. {\bf A107} (1985)
21-25;  ``Low-Energy
Scattering of Nonabelian Magnetic Monopoles'',
Phil.Trans.Roy.Soc.Lond. {\bf A315} (1985)  459-469.}%
\atiyahhitch\ and that the combined system of $N$ D6-branes on top of
the O6-plane is described locally by the $D_{N}$ ALE singularity. This
is the orbifold singularity ${\bf C}^2/{\bf {\hat D}}_{N-2}$, where
${\bf {\hat D}}_{N-2}$ is the Dihedral group with a ${\bf Z}_2$ central 
extension \foot{This discrete
subgroup of $SU(2)$ 
has order $4N-8$ and has two generators  $a$ and $b$
which satisfy $a^{2N-4}=b^4=1$ and $ba=a^{-1}b$. The action of ${\bf {\hat
D}}_{N-2}$ on ${\bf C}^2$ can be found, for example, in Table 1 of
\nref\aspinwall{P.S. Aspinwall, ``K3 Surfaces and String Duality'',
hep-th/9611137.}%
\aspinwall .}. The $SO(2N)$ gauge symmetry
arises in the M-theory description from membranes wrapping the
shrunken two-cycles at 
the orbifold singularity.

To summarize, $N$ coincident D6-branes in flat space are described in eleven
dimensions by an $A_{N-1}$ ALE singularity and $N$ D6-branes on top of an
O6$^-$-plane by a $D_N$ ALE singularity. The corresponding orbifold groups
act on the normal direction of the D6-branes and on the M-theory
circle. In the rest of the paper we will explore the M-theory
description of D6-branes  and D6-branes on top of an O6-plane wrapping non-trivial supersymmetric cycles in
various non-trivial backgrounds.

\newsec{General Strategy}

We want to consider Type IIA supersymmetric configurations of
D6-branes wrapping cycles in spaces with reduced holonomy and to find the
corresponding M-theory description. In order for
the D6-branes to preserve some of the supersymmetry left unbroken by the
compactification space, the cycle which is wrapped must be
supersymmetric
\nref\cali{K. Becker, M. Becker and A. Strominger, ``Fivebranes,
Membranes and Non-Perturbative String Theory'', Nucl.Phys. {\bf B456}
(1995) 130;
H. Ooguri, Y. Oz and Z. Yin, ``D-Branes on Calabi-Yau Spaces and Their
Mirrors'', Nucl.Phys. {\bf B477} (1996) 407;
 K. Becker, M. Becker, D.R. Morrison, H. Ooguri, Y. Oz and
Z.Yin, ``Supersymmetric Cycles in Exceptional Holonomy Manifolds and
Calabi-Yau 4-Folds'', Nucl.Phys. {\bf B480} (1996) 225.}
\nref\topo{ M. Bershadsky, V. Sadov and C. Vafa, ``D-Branes and
Topological Field Theories'', Nucl.Phys. {\bf B463} (1996) 420.}%
\refs{\cali,\topo}.  The BPS condition on a
supersymmetric cycle on a manifold with holonomy group $G$ can be
written in terms of the corresponding $G$-structure. 
The
following table summarizes the classification of supersymmetric cycles
(table 3 in 
\nref\nonlin{M. Mari\~no, R. Minasian, G. Moore and A. Strominger,
``Nonlinear Instantons from Supersymmetric p-Branes'', JHEP {\bf 0001}
(2000) 005.}%
\nonlin ):
$$\vbox{\offinterlineskip\halign{ \strut # height10pt depth 13pt&
\quad#\quad\hfill\vrule&\quad$#$\quad\hfill\vrule&
\quad$#$\quad\hfill\vrule& \quad$#$\quad\hfill\vrule&
\quad$#$\quad\hfill\vrule& \quad$#$\quad\hfill\vrule\cr
\noalign{\hrule} \vrule& p+1& SU(2)&SU(3)&G_2&SU(4)&\hbox{Spin}(7)\cr
\noalign{\hrule} \vrule& 2& {\rm divisor/SLag}&{\rm
holomorphic}&-&{\rm holomorphic}&-\cr \noalign{\hrule} \vrule& 3&
-&{\rm SLag}&{\rm associative}&-&-\cr \noalign{\hrule} \vrule& 4&
X&{\rm divisor}&{\rm coassociative}&{\rm Cayley}&{\rm Cayley}\cr
\noalign{\hrule} \vrule& 5& -&-&-&-&-\cr \noalign{\hrule} \vrule&
6& -&X&-&{\rm divisor}&-\cr \noalign{\hrule} \vrule&7&
-&-&X&-&-\cr \noalign{\hrule} \vrule& 8& -&-&-&X&X\cr
\noalign{\hrule} }}$$
\centerline{Table 1. Supersymmetric Cycles in irreducible holonomy manifolds.}
\bigskip

 All the above cycles preserve
one-half of supersymmetry except the Cayley cycles of a Calabi-Yau
four-fold which instead
preserve one quarter of the supersymmetries. Therefore,  the
supersymmetric Type IIA
backgrounds  with                wrapped D6-branes which can be
described by M-theory on a  manifold with irreducible  holonomy are severely
constrained by supersymmetry.

\noindent
There are essentially two types of situations that need to be
considered:

1) The D6-branes wrap a supersymmetric
cycle in such a way that the D6-branes fill completely the space
transverse to the IIA compactification manifold. The field theory
one obtains  lives in the transverse Minkowski space.
 Then the local M-theory
description is given by a manifold one dimensional higher in the list
of manifolds with irreducible holonomy. In this fashion one can find examples of the following
lifts of holonomy groups (IIA$\rightarrow$ M): $SU(3)\rightarrow G_2$
and $G_2\rightarrow 
\hbox{Spin}(7)$. 

2)The D6-branes wrap a supersymmetric
cycle in such a way that the D6-branes are codimension one in the 
transverse Minkowski space to the IIA compactification manifold. The
field theory one 
obtains lives in codimension one on the transverse space.
Then the local M-theory
description -- near the D6-branes -- is given by a manifold
of two dimensions higher in the list of manifolds with
irreducible  holonomy. In this fashion one can find examples of the following
lifts of holonomy groups (IIA$\rightarrow$ M): $SU(2)\rightarrow SU(3)$
and $SU(3)\rightarrow SU(4)$.

The effective gauge theory living on the D-branes is a topological field
theory \topo. This topological field theory is a gauge theory in which 
the scalars parametrizing the positions of the D-branes
are twisted and transform as sections of the normal
bundle\foot{Recently, supergravity duals of these topological field
theories have been constructed, see e.g 
\nref\sugratopo{J. Maldacena and C. N\'u\~nez, ``Supergravity description
of field theories on curved manifolds and a no go theorem'',
hep-th/0007018; J. Maldacena and C. Nu\~nez, ``Towards the large N
limit of pure N=1 super Yang Mills'', 
Phys.Rev.Lett. {\bf 86} (2001) 588; B.S. Acharya, J.P. Gauntlett and
N. Kim, ``Fivebranes Wrapped On Associative Three-Cycles'',
hep-th/0011190; J.P. Gauntlett,
N. Kim and D. Waldram, ``M-Fivebranes Wrapped on Supersymmetric
Cycles'', hep-th/0012195; H. Niemer and Y. Oz, ``Supergravity and
D-branes Wrapping Supersymmetric 3-Cycles'', hep-th/0011288;
C. N\'u\~nez, L.Y. Park, M. Schvellinger and T.A. Tran, ``Supergravity
duals of gauge theories from F(4) gauged supergravity in six
dimensions'', hep-th/0103080.}%
\sugratopo .}.
Therefore, given a supersymmetric cycle one can construct the
supersymmetric gauge
theory by analyzing the normal bundle of the
supersymmetric cycle. In this paper we will concentrate on
supersymmetric cycles which are {\it rigid} so that there are no
scalars describing 
the possible deformations of the cycle inside the curved manifold.
 When the D6-branes are codimension one, we get a
massless real scalar field  which is part of the vector multiplet. It 
describes the positions of the D6-branes
in the transverse direction. The M-theory realization of these
backgrounds with 
D6-branes will geometrically engineer these gauge theories.

The local geometry of the wrapped N D6-branes in M-theory is as
follows. Near the location of the D6-branes, the D6-branes are
represented in M-theory by the $A_{N-1}$ ALE singularity. Since some
of the transverse directions of the D6-branes are curved and
non-trivially fibered over the supersymmetric cycle, the lift to
eleven dimensions is just given by the $A_{N-1}$ ALE singularity
fibered over the supersymmetric cycle. As we will show in the sections
that follow we will be able to recognize such geometries as particular
manifolds with irreducible holonomy.

We can also consider the situation in which we mod out the geometry on
which the D6-branes are embedded by an involution $\sigma$. We will
require that the involution have as fixed set the supersymmetric cycle
on which the D-branes are wrapped and that the fixed set satisfies the
corresponding supersymmetric cycle condition.
 In order to obtain
a supersymmetric configuration this involution must be accompanied by
an orientifold projection. The appropriate orientifold group is
given\foot{In the situation where the D6-branes are codimension one
in the transverse space, one must also reflect along the transverse
coordinate.} 
by $G=\{1,\Omega (-1)^{F_L}\sigma\}$. Therefore, by performing this
orientifold projection, one obtains a curved orientifold six-plane (O6-plane)
with $N$ D-branes on top of it. Following the same argument we used
to uplift the description of the curved D6-branes to M-theory we see
that the M-theory description of the curved D6+O6 configuration is
given by the $D_N$ ALE singularity fibered over the corresponding
supersymmetric cycle. In the examples we will consider we will
identify this fibration with a certain reduced holonomy manifold.

In the following sections we will provide concrete examples of these
lifts.

\newsec{From $SU(2)$ holonomy to $SU(3)$ holonomy via D6-Branes }

We lift various IIA configurations and recover the
Calabi-Yau three-folds used, for example,  in \geom\ to geometrically
engineer four 
dimensional ${\cal N}=2$ gauge theories.

The simplest\foot{Lifting to eleven dimensions a configuration of D6-branes 
wrapped on the entire K3 follows trivially from the discussion on the previous
section. The corresponding eleven dimensional geometry is just
${\bf R}^{1,2}\times$Taub-NUT$\times$ K3.} case to consider is the
M-theory description of N D6-branes 
wrapping the  ${\bf S}^2$ in  $T^*{\bf S}^2$, the cotangent bundle of
${\bf S}^2$. This 
space appears when resolving 
or deforming an $A_1$ 
singularity of K3. For instance, $T^*{\bf S}^2$ is described by
\eqn\deform{
z_1^2+z_2^2+z_3^2=r.}
If we rewrite $z_j=a_j+ib_j$ for $j=1,2,3$ and take $r$ to be real and
positive,
the real and imaginary parts of \deform\ lead to
$\vec{a}^{\;2}-\vec{b}^{\;2}=r$ and $\vec{a}\cdot \vec{b}=0$. Since
$\vec{u}\cdot \vec{u}=1$, where 
 $\vec{u}={\vec{a}/ \sqrt{\vec{b}^{\;2}+r}}$, ${\vec u}$  generates an
${\bf S}^2$. Moreover, since $ \vec{b}\cdot
\vec{u}=0$, $b_i$ spans the cotangent directions of ${\bf S}^2$, so 
\deform\ describes the cotangent bundle of ${\bf S}^2$.
This space is topologically  ${\bf R}^2\times {\bf S}^2$ and admits a hyper-Kahler
metric with $SU(2)$ holonomy. The ${\bf S}^2$ is a holomorphic cycle and branes
wrapping it preserve one-half of the supersymmetries left unbroken by
the geometry.

We now consider N D6-branes wrapped on the supersymmetric ${\bf S}^2$  inside
${\bf R}^{1,5}\times T^*{\bf S}^2$ and stretched along the $x^1\ldots x^4$
directions.  
Since the normal bundle of ${\bf S}^2$ $N({\bf S}^2)$ is trivial, the
field theory 
living on the branes is a supersymmetric five-dimensional pure
$SU(N)$ gauge theory with eight real supercharges\foot{The vector
multiplet has a real scalar so there is a non-trivial Coulomb
branch whose prepotential is constrained by gauge invariance to be
cubic in the vector superfields
\nref\seiberg{N. Seiberg, ``Five Dimensional SUSY Field Theories,
Non-trivial Fixed Points and String Dynamics'', Phys.Lett. {\bf B388}
(1996) 753.}%
\seiberg .}. Alternatively, one
can simply derive
\nref\quiver{M.R. Douglas and G. Moore, ``D-branes, Quivers, and ALE
Instantons'', hep-th/9603167; C.V. Johnson and R.C. Myers, ``Aspects
of Type IIB Theory on ALE Spaces'', Phys.Rev. {\bf D55} (1997) 6382.}%
\quiver\ this gauge theory  by representing the wrapped
D6-branes by $N$ fractional
\nref\fract{
J. Polchinski, ``Tensors from K3 Orientifolds'', Phys. Rev. {\bf D55}
(1997) 6423; M.R. Douglas, ``Enhanced Gauge Symmetry in M(atrix)
Theory'', JHEP {\bf 9707} (1997) 004; D.E. Diaconescu,
M.R. Douglas, 
and J. Gomis, ``Fractional Branes and Wrapped Branes'', JHEP {\bf
9802} (1998) 013.}%
\fract\  D4-branes at a ${\bf C}^2/{\bf Z}_2$ orbifold
singularity.

Now, in eleven dimensions we must obtain a geometrical background which leads
to such a five-dimensional gauge theory.  As explained in the previous
section the geometry we get is that of an $A_{N-1}$ ALE singularity fibered
over the ${\bf S}^2$. This geometry is indeed a Calabi-Yau three-fold \geom\
and
gives rise to a five dimensional $SU(N)$ gauge theory. The exact
prepotential in the Coulomb branch of this gauge theory
was computed from M-theory 
on this three-fold geometry in 
\nref\dege{K. Intriligator, D.R. Morrison and N. Seiberg,
``Five-Dimensional Supersymmetric Gauge Theories and Degenerations of
Calabi-Yau 
   Spaces'', Nucl.Phys. {\bf B497} (1997) 56.}%
\dege . The computation reduces to evaluating the classical
intersection numbers of the divisors one obtains when resolving the
 curve of
 $A_{N-1}$ singularities which when blown up break the $SU(N)$ gauge group to its
 maximal torus.

In the Type IIA description one can perform an orientifold projection combined
with the following involution
\eqn\invol{
{\sigma}:\Big\{ 
\matrix{ z_i & \rightarrow & {\bar z_i}& \quad i=1,2,3\cr
x_5 & \rightarrow & -x_5& \cr}.}
   This involution keeps fixed the real part of equation \deform\ fixed
which describes the supersymmetric ${\bf S}^2$. Therefore, we obtain
an O6-plane wrapping the ${\bf S}^2$. Then, the gauge theory
 on the D6-branes is a five-dimensional
$SO(2N)$ gauge theory with eight real supercharges. 

It is simple to incorporate in the M-theory  description of this
background the
effect of the O6-plane. The geometry that one gets is that of a $D_N$
ALE singularity fibered over the base ${\bf S}^2$. This Calabi-Yau
appears  in \geom\ and M-theory on it gives a
supersymmetric $SO(2N)$ gauge theory. By analysing the resolution
of this singularity reference \dege\ reproduced the exact prepotential in the
Coulomb branch of the gauge theory from  classical geometry\foot{Both
the $SU(N)$ and $SO(2N)$ pure gauge theories have a convex prepotential along the Coulomb
branch and have a non-trivial fixed point of the renormalization
group at the origin \seiberg\geom\ .}.

It is natural to ask for the M-theory description of wrapped D6-branes
when the D6-branes wrap a collection of ${\bf S}^2$'s of a local K3
geometry. We will consider the well known
ADE ALE spaces. The basic geometry is that of a collection of $r$
${\bf S}^2$'s whose intersection form is $I_{ab}=-C_{ab}$ where
$a,b=1,\ldots ,r$,
which is  (minus) the Cartan matrix
of the corresponding ADE algebra of rank $r$. One can consider
wrapping $N_a$ D6-branes on the $a$-th ${\bf S}^2$ for $a=1,\ldots
,r$. In order to obtain a supersymmetric gauge theory on this
collection of D6-branes, one must ensure that all the ${\bf S}^2$'s are
holomorphic (supersymmetric) with respect to the same complex
structure\foot{For the $A_r$ ALE spaces, for which an explicit metric
is known \eguchihan  , one can show that the $r$ homology generators
are holomorphic with respect to the same complex structure when the
$r$ vectors in ${\bf R}^3$ -- from which one constructs the ${\bf S}^2$'s
by fibering these vectors with the $U(1)$ fiber -- are
collinear.}. Then the gauge theory one obtains is a five dimensional
theory with eight real supercharges with gauge group
\eqn\gauge{
G=\prod_{a=1}^r SU(N_a)}
and the following hypermultiplet  content 
\eqn\hyper{
{1\over 2}\oplus_{a\neq b} I_{ab} (N_a,{\bar N_b}).}
This gauge theory can be easily derived by representing the wrapped
D6-branes as  fractional D4-branes probing the ${\bf C}^2/{\bf
\Gamma}$ orbifold singularity, 
where ${\bf \Gamma}$ is the discrete subgroup of $SU(2)$ whose
representation theory can be associated with the extended ADE Dynkin
diagram\foot{$A_r={\bf Z}_{r+1}, D_r={\bf {\hat D}}_{r-2}$; $E_6,E_7$
and $E_8$ correspond respectively to the ${\bf Z}_2$ centrally
extended tetrahedral, octohedral and isocahedral groups.}.

The M-theory description of this Type IIA background is as
follows. One has a fibration structure whose base is given by a
chain of $r$ ${\bf S}^2$'s which intersect according to  a particular
ADE Dynkin diagram of rank $r$. Which ADE diagram appears is determined by
which ADE ALE space is used in the Type IIA description. On the $a$-th
${\bf S}^2$ in the base one has a $A_{N_a-1}$ ALE singularity fibered
over it so the total geometry is that of a chain of ${\bf S}^2$'s
intersecting according to the ADE Dynkin diagram and on each  ${\bf
S}^2$ there is an $A$-type  singularity fibered over it. The type of
singularity on a particular ${\bf S}^2$ in the base is determined by the number of
D6-branes that wrap that cycle in the Type IIA description. This
geometry is a Calabi-Yau three-fold and M-theory on it
results in the required quiver gauge theory \foot{The matter appears as usual
from loci of enhanced symmetry in the base.} and reproduces the cubic
prepotential along the Coulomb branch \geom .

We conclude this section by identifying the Type IIA background which
lifts to M-theory on a family of Calabi-Yau manifolds considered in
\geom . In \geom \ it was shown that one can generalize the base
geometry that we have just discussed such that the base ${\bf S}^2$'s 
intersect according to Dynkin diagram of the affine ${\hat A}{\hat
D}{\hat E}$ groups. The hypermultiplets are determined like in
\hyper\ but now $C_{ab}$ is the Cartan matrix of ${\hat A}{\hat
D}{\hat E}$.
It is natural to try to identify the Type IIA
geometry with $SU(2)$ holonomy which lifts, in the presence of
D6-branes, to this family of Calabi-Yau manifolds.

When the  base ${\bf S}^2$'s intersect according to the ${\hat
A}_r$ Dynkin diagram, the corresponding Type IIA geometry is 
the deformation of the Type I$_{r+1}$ fibre singularity of
an elliptically fibered K3.
 This K3  is described by the Weierstrass form 
\eqn\fib{
y^2=x^3+a(z)x+b(z),}
where $(x,y)$ are affine coordinates on ${\bf \P}^2$ and $z$ is the
coordinate on the base of the elliptic fibration parametrizing a copy
of ${\bf C}$. The Type I$_{r+1}$ singularity appears when the
discriminant of \fib\ has a zero of order $r+1$ at $z=0$ and both $a(z)$
and $b(z)$ don't have a zero at $z=0$. The deformation of this
singularity generates a chain of  $r+1$ ${\bf S}^2$'s which intersect
according to the ${\hat
A}_r$ Dynkin diagram. Wrapping D6-branes on these two-cycles
lifts to eleven dimensions to the Calabi-Yau three-fold we were
looking for. One can find in a similar fashion the Type IIA
realization when the ${\bf S}^2$'s in the base of the three-fold have
a different affine intersection form. This corresponds to studying
Type IIA with D6-branes on the deformation of certain fiber
singularities -- see e.g \aspinwall\ --
of an
elliptically fibered K3.

\newsec{From $SU(3)$ holonomy to $G_2$ holonomy via D6-Branes }

For completeness, we briefly summarize an example of this lift
which appeared recently in \acha\atmalva . The Calabi-Yau three-fold
they considered is $T^*{\bf S}^3$, the cotangent bundle of ${\bf
S}^3$. This manifold appears as the deformation of the quadric on
${\bf C}^4$
\eqn\quadric{
z_1^2+z_2^2+z_3^2+z_4^2=r.}
In a similar fashion as in \deform , one may show that for real $r$ \quadric\
describes $T^*{\bf
S}^3$. One can now wrap N D6-branes branes over the ${\bf
S}^3$, which is a Special Lagrangian submanifold. The branes
completelly fill the transverse ${\bf R}^{1,3}$ Minkowski space and
give rise a four dimensional $SU(N)$ gauge theory with four
supercharges. 

In \acha\atmalva\  the local description of this background was
described as M-theory on a certain space with $G_2$ holonomy. The
D6-branes lead to an $A_{N-1}$ ALE singularity which is fibered over
the ${\bf
S}^3$. This geometry can be understood as a ${\bf Z}_N$ orbifold of an
${\bf R}^4$ fibration over ${\bf
S}^3$. Supersymmetry dictates that this space must have $G_2$
holonomy. It is known that the spin bundle of ${\bf
S}^3$ $S({\bf
S}^3)$, whose fibers are topologically ${\bf R}^4$, admits a metric
with $G_2$ holonomy
\nref\metrics{R.Bryant and S. Salomon, ``On the construction of some
complete metrics with exceptional holonomy'', Duke Math. J. {\bf 58}
 (1989) 829; G.W. Gibbons, D.N. Page and C.N. Pope, ``Einstein Metrics
on ${\bf S}^3$,${\bf R}^3$ and ${\bf R}^4$ Bundles'',
Commun.Math.Phys. {\bf 127} (1990) 529.}%
\metrics . Moreover, the $G_2$ structure on this space
contains an $SU(2)^3$ symmetry group. Therefore, modding out by any
discrete subgroup of $SU(2)^3$ will preserve the $G_2$
structure. Indeed, it is easy to show that embedding ${\bf Z}_N$ in
one of the  $SU(2)'s$ acts on the fibers precisely as we expect from the lift
of Type IIA D6-branes on $T^*({\bf
S}^3)$. Therefore, the Type IIA configuration is represented in eleven
dimensions by a ${\bf Z}_N$ orbifold of $S({\bf
S}^3)$ which has a $G_2$ structure.

Sinha and Vafa 
\sinvafa\
considered the Type IIA orientifold background obtained by modding out by the
following involution
\eqn\invol{
\sigma: z_i\rightarrow {\bar z}_i \qquad i=1,\ldots,4.}
This involution gives rise to an O6-plane wrappping the ${\bf
S}^3$, which is left fixed by \invol . The gauge theory on the
D6-branes on top of this O6-plane is a four dimensional  ${\cal N}=1$ 
$SO(2N)$ gauge theory.

From the general discussion in section $3$ such background is
described in eleven dimensions by a $D_{N}$ ALE singularity fibered
over ${\bf
S}^3$. This is obtained by having the discrete group ${\bf {\hat D}}_{N-2}$
act on the ${\bf R}^4$ fibers. Since ${\bf {\hat D}}_{N-2}\in SU(2)$, one
can mod
out $S({\bf S}^3)$ by ${\bf {\hat D}}_{N-2}$ by appropiately embedding
${\bf {\hat
D}}_{N-2}$ in one 
of the $SU(2)$ symmetry groups of $S({\bf S}^3)$ such that the
quotient space has a $G_2$ structure and ${\bf D}_{N-2}$ acts as
required on the fibers.

Various aspects of this lift and generalizations to other
three-folds have recently appeared in
\refs{\sinvafa,\achar,\achavafa,\cacintrivafa}.

\newsec{From $SU(3)$ holonomy to $SU(4)$ holonomy via D6-Branes }

In order to accomplish this lift the D6-branes must be codimension one
in the transverse ${\bf R}^{1,3}$ Minkowski space. Therefore, the
D6-branes must wrap a divisor in the Calabi-Yau three-fold. In this
section we will provide a simple example of this lift.

Consider the non-compact Calabi-Yau geometry described by the
${\cal O}(-3)$ bundle
over ${\bf \P}^2$. The string sigma model  on this three-fold has a simple
description in 
terms of the linear sigma model approach of
\nref\witten{E. Witten, ''Phases of $N=2$ Theories In Two
Dimensions'', Nucl.Phys. {\bf B403} (1993) 159.}%
\witten . We analyze the vacuum structure of a two dimensional ${\cal
N}=(2,2)$ $U(1)$ gauge theory with four chiral superfields
$(z_1,z_2,z_3,z_4)$ which carry charges $(1,1,1,-3)$ under  the gauge
group. In the presence of a Fayet-Iliopoulos term $r$, the solution of
the D-term equation of this model is described by
\eqn\vac{
M:\qquad |z_1|^2+ |z_2|^2+ |z_3|^2-3 |z_4|^2=r.}
The vacuum of the theory is $M/U(1)$, where the $U(1)$ action is
specified by the $U(1)$ charges of $z_i$.

The phase $r>0$ describes an exceptional ${\bf \P}^2$ 
parametrized by coordinates $z_1,z_2,z_3$ which due to \vac\ cannot
vanish simultaneously. The  coordinate  $z_4$ describes a complex line
fibered over ${\bf \P}^2$. The vacuum manifold $M/U(1)$
can be identified\foot{This Calabi-Yau is the crepant resolution of
the ${\bf C}^3/{\bf Z}_3$ orbifold singularity which appears in the
linear sigma model approach in the phase when $r<0$; then  $z_4$
cannot vanish and the exceptional divisor is blown down. More details
on the moduli space of this model can be found in
\nref\bound{D.E. Diaconescu and J. Gomis, ``Fractional Branes and
Boundary States in Orbifold Theories'', JHEP {\bf 0010} (2000) 001.}%
\bound .} with the ${\cal O}(-3)$ bundle
over ${\bf \P}^2$. The exceptional ${\bf \P}^2$ is a holomorphic cycle
so branes wrapping it are supersymmetric.

We consider $N$ D6-branes wrapping the ${\bf \P}^2$. The field
theory on the branes is a three-dimensional ${\cal N}=2$ pure $SU(N)$
gauge theory\foot{There is a subtlety in this example. Since ${\bf
\P}^2$ is not a Spin manifold we must turn on a half-integral flux on
the D-brane in order to avoid global anomalies
\nref\anom{D.S. Freed and E. Witten, ``Anomalies in String Theory with
D-Branes'', hep-th/9907189.}%
\anom . We will
return to this issue in the next section.}. Classically, this theory
has a Coulomb branch 
parametrized by $N-1$ complex scalars which are formed from the real
scalars in the vector multiplet together with the real scalars one
gets by dualizing the photons one gets along the Coulomb
branch. Semiclassically, Yang-Mills instantons generate a
superpotential for the complex scalars
\nref\ahwit{I.A. Affleck, J.A. Harvey and E. Witten, ``Instantons and
(Super)Symmetry Breaking in (2+1)-Dimensions'', Nucl.Phys. {\bf B206}
(1982) 413.}%
\ahwit\ break supersymmetry.

 The local M-theory description of this background
is given by an $A_{N-1}$ ALE singularity fibered over  ${\bf \P}^2$
which is a Calabi-Yau four-fold. Analyzing the zero mode spectrum leads
to a ${\cal N}=2$ three-dimensional pure $SU(N)$ gauge
theory\foot{If the complex
surface $Z$ over which the singular ALE sits has non-zero Betti
numbers $h^{1,0}(Z)$ and $h^{2,0}(Z)$ one gets chiral
multiplets
\nref\katzvafa{S. Katz and C. Vafa, ``Geometric Engineering of N=1
Quantum Field Theories'', Nucl.Phys. {\bf B497} (1997) 196.}%
\katzvafa . Since $h^{1,0}({\bf \P}^2)=h^{2,0}({\bf \P}^2)=0$ we get
pure gauge theory.}.
Precisely the four-fold geometry we have
encountered can be used to exactly reproduce the superpotential along
the Coulomb branch of the gauge theory
\katzvafa . In the geometrical picture going to the Coulomb branch
corresponds to resolving the singularities on the fiber which lead in
the total geometry to a collection of divisors. The vevs of the
complex scalars in the Coulomb branch correspond to the complexified
blow up parameters. The superpotential
then arises, when  certain topological conditions 
\nref\witsup{E. Witten, ``Non-Perturbative Superpotentials In String
Theory'', Nucl.Phys. {\bf B474} (1996) 343.}%
\witsup\ are satisfied, by wrapping Euclidean five-branes over the
divisors. It was shown in \katzvafa\ that the divisors one obtains by
derforming the singularity we discussed reproduces the expected field
theory superpotential of the three dimensional $SU(N)$ gauge theory.

We can orientifold the Type IIA background by modding out by the
following involution
\eqn\invol{
{\sigma}:\Big\{ 
\matrix{ z_4 & \rightarrow & -z_4& \cr
x_3 & \rightarrow & -x_3& \cr}.}
The action of $\sigma$ on the three-fold \vac\ leaves the ${\bf \P}^2$
fixed and gives rise to a curved orientifold plane. Therefore,
 we can have $N$ D6-branes on top of the O6-plane wrapping
the  ${\bf \P}^2$ inside the ${\cal O}(-3)$ bundle
over ${\bf \P}^2$. Thus, the gauge theory on the D6-branes is a three
dimensional ${\cal N}=2$ pure $SO(2N)$ gauge theory. The local
M-theory description of this Type IIA system is described by the
Calabi-Yau four-fold given by a $D_N$ ALE singularity fibered over
${\bf \P}^2$. This geometry was used in \katzvafa\ to reproduce the
superpotential on the Coulomb branch of the gauge theory using
five-brane instantons.

\newsec{From $G_2$ holonomy to $\hbox{Spin}(7)$ holonomy via D6-Branes }

In order to accomplish this lift the D6-branes must fill the
transverse ${\bf R}^{1,2}$ Minkowski space. We must then look for
manifolds with $G_2$ holonomy which have a coassociative four
cycle. In order to avoid complications with tadpoles we will consider
non-compact geometries. 

In the
literature \metrics , there are only three complete metrics 
on  seven manifolds which admit a
$G_2$ structure\foot{A $G_2$ structure is a globally defined three-form
$\Phi$ which is covariantly constant, closed, co-closed and
invariant under the group 
$G_2$. $G_2$ is the subgroup of $SO(7)$ which leaves the
multiplication table of imaginary octonions invariant.}. Of these
geometries, two have a coassociative four cycle\foot{The other metric
is a $G_2$ holonomy metric on the spin bundle over ${\bf S}^3$ $S({\bf
S}^3)$ which
appeared in the previous section. This geometry has an associative
three-cycle.}. They are metrics on the bundle of anti-self-dual
two-forms over the four-manifold ${\bf Z}$ $\Sigma({\bf Z})$, where
${\bf Z}={\bf S}^4$ or ${\bf Z}={\bf {\P}}^2$. Topologically, the
fibers  are ${\bf R}^3$ and ${\bf Z}$ is a supersymmetric
four-cycle. Therefore, wrapping a collection of D6-branes over ${\bf
Z}$ gives rise
to a three-dimensional ${\cal N}=1$ gauge theory. Since the normal
bundle of ${\bf Z}$ is trivial\foot{The normal bundle is the bundle of
anti-self-dual two-forms. Since $\hbox{dim}(H^2_-({\bf Z})=0)$, there
are no scalars.} one obtains a pure $SU(N)$ gauge
theory. Theories with three-dimensional ${\cal N}=1$ supersymmetry do
not have the familiar constraints of holomorphy of four dimensional
${\cal N}=1$ field theories, therefore one cannot make exact statements
about their non-perturbative dynamics.

We will start by identifying the M-theory description when the
D6-branes wrap the ${\bf S}^4$ in $\Sigma({\bf S}^4)$. The M-theory
geometry is given by the $A_{N-1}$ ALE singularity fibered over ${\bf
S}^4$ or equivalently by an ${\bf R}^4$ bundle over ${\bf
S}^4$ where the cyclic group ${\bf Z}_N$ acts on the  ${\bf R}^4$
fibers. Supersymmetry requires that this space admit a
$\hbox{Spin}(7)$ structure\foot{A $\hbox{Spin}(7)$ structure is a
globally defined self-dual four-form 
$\Omega$ which is covariantly constant, closed and
invariant under the group 
$\hbox{Spin}(7)$.}. Fortunately, we can identify this space with 
an orbifold of the known \metrics\ eight-dimensional space which
admits a complete  metric
with  $\hbox{Spin}(7)$ holonomy. This space is the spin
bundle over ${\bf
S}^4$ $S({\bf
S}^4)$, whose fibers are topologically ${\bf R}^4$ and admits a
$\hbox{Spin}(7)$ structure with an $SU(2)\times Sp(2)$ symmetry. One
can now embedd ${\bf Z}_N$ on $SU(2)$ while preserving the
$\hbox{Spin}(7)$ structure. Moreover, this embedding acts
geometrically by natural action of ${\bf Z}_N$ on the ${\bf R}^4$
fibers. Therefore, we have identified the M-theory description of
D6-branes wrapping the ${\bf S}^4$ in $\Sigma({\bf S}^4)$ as an
abelian orbifold of $S({\bf
S}^4)$ which preserves the $\hbox{Spin}(7)$ structure.

One can also consider the case when the D6-branes wrap ${\bf
\P}^2$ in $\Sigma({\bf
\P}^2)$. Since ${\bf
\P}^2$ is not a spin manifold, the field strength $F$ of the $U(N)$
``gauge field'' on the D6-branes\foot{More precisely $F$ is not a
 connection on a vector bundle but rather a Spin$^c$ structure.}
 does not obey conventional Dirac
quantization
\anom . One has 
\eqn\Dirac{
\int_{{\bf \P}^1}{F\over 2\pi}=n+{1\over 2},}
where $n$ is an integer and ${\bf \P}^1$ generates $H_2({\bf
\P}^2,{\bf Z})$. In particular, the flux cannot vanish. Despite this
flux, the effective three-dimensional  theory one obtains on the
branes is also
a pure $SU(N)$ gauge theory.

The eleven dimensional description of this Type IIA background is
somewhat subtle. Naively, one gets near the location of the
D6-branes an $A_{N-1}$ ALE singularity fibered over ${\bf \P}^2$. 
Supersymmetry
suggest, in analogy with the previous example, that it must be
possible to put a metric on an ${\bf R}^4$ bundle over ${\bf \P}^2$
which is complete and has $\hbox{Spin}(7)$ holonomy. Moreover, such
$\hbox{Spin}(7)$ structure must have  an $SU(2)$ symmetry on which one can
embedd ${\bf Z}_N$ in such a way that it acts on the ${\bf R}^4$
fibers in the usual fashion. We have not found in the literature a
metric such an ${\bf R}^4$ bundle over ${\bf \P}^2$ with
$\hbox{Spin}(7)$ holonomy but duality suggest that it must exist. 

Just like in previous sections it is possible to consider the M-theory
description when there is  an orientifold six-plane wrapping ${\bf
Z}$. The appropiate involution is given by acting by ${\bf -1}$ on the
fibers such that ${\bf
Z}$ is fixed by the involution. One then gets a three-dimensional ${\cal N}=1$ $SO(2N)$
gauge theory on the D-branes. The corresponding M-theory geometry is
given by a $D_N$ ALE singularity fibered over  ${\bf
Z}$ which admits a $\hbox{Spin}(7)$ structure.

The analysis of lifts to M-theory of wrapped D6-branes on supersymmetric cyles
which are not Spin  raises an interesting
question. As we have remarked, absence of global anomalies forces
\anom ,
even for a single wrapped D6-brane over ${\bf \P}^2$, to turn on a
half-integral flux \Dirac . In the absence of such a flux one expects
the M-theory description to be given by Taub-NUT space. In the
presence of this flux, it is not clear what must be the appropiate
modification to the M-theory solution.
It would be interesting to identify\foot{This constraint
was derived
\nref\dualrel{E. Witten, ``Duality Relations Among Topological Effects
In String Theory'', JHEP {\bf 0005} (2000) 031.}%
\dualrel\  from an M-theory perspective when one considers instead
D4-branes. Then such a condition can be derived from an eleven
dimensional viewpoint by demanding consistency of the M-theory five-brane
partition function.} the M-theory origin of the
constraint on Dirac quantization whenever a D6-brane wraps a
manifold which is not spin but  is Spin$^c$.


\vskip 1cm
{\bf Acknowledgements:}
\vskip 1cm

We would like to thank M. Atiyah, A. Brandhuber, R. Corrado, E. Diaconescu,
J. Gauntlett, S. Gukov, L. Motl, H. Ooguri and 
E. Witten  for very useful discussions. Part of this work was done
during the M-theory workshop at the ITP. We would like to thank the
organizers for providing a stimulating atmosphere. This research was
supported in part by the National Science Foundation under Grant
No. PHY99-07949 and by the  DOE under grant no. DE-FG03-92-ER
40701.

\listrefs

\end